\documentclass[prl,superscriptaddress,onecolumnpage,notitlepage]{revtex4-1}

\pdfoutput=1
\usepackage{graphicx} 
\usepackage{tikz}
\usepackage{amsmath}
\usepackage{amssymb}
\usepackage{natbib}
\usepackage{color}
\usepackage{float}
\usepackage{epstopdf}
\usepackage[normalem]{ulem}
\usepackage{pgfplots,pgfplotstable}
\usepackage{soul}
\usepackage{accents}
\usepackage{nameref}
\usepackage{balance}
\usepackage[all]{nowidow}
\usepackage{wrapfig}

\newcommand{\rpm}{\raisebox{.2ex}{$\scriptstyle\pm$}}
 
\begin{document}

\title{Buckling of geometrically confined shells}
\author{Lucia Stein-Montalvo}
\email{lsmontal@bu.edu}
\affiliation{
Department of Mechanical Engineering, Boston University, Boston, MA, 02215.
}%

\author{Paul Costa}
\affiliation{Mechanical Engineering, \`Ecole Polytechnique, 91128 Palaiseau, France.
}%

\author{Matteo Pezzulla}
\affiliation{
	Department of Mechanical Engineering, Boston University, Boston, MA, 02215.
}%
\affiliation{Institute of Mechanical Engineering, \`Ecole polytechnique f\`{e}d\`{e}rale de Lausanne, 1015 Lausanne, Switzerland.
}%

\author{Douglas P. Holmes}
\email{dpholmes@bu.edu}
\affiliation{
Department of Mechanical Engineering, Boston University, Boston, MA, 02215.
}%

\date{\today}

\begin{abstract}
We study the periodic buckling patterns that emerge when elastic shells are subjected to geometric confinement. \textit{Residual swelling} provides access to range of shapes (saddles, rolled sheets, cylinders, and spherical sections) which vary in their extrinsic and intrinsic curvatures. Our experimental and numerical data show that when these structures are radially confined, a single geometric parameter -- the ratio of the total shell radius to the amount of unconstrained material -- predicts the number of lobes formed. We then generalize our model to account for a relaxation of the in-plane constraint by interpreting this parameter as an effective foundation stiffness. Experimentally, we show that reducing the transverse confinement of saddles causes the lobe number to decrease according to our effective stiffness model. Hence, one geometric parameter captures the wave number through a wide range of radial and transverse confinement, connecting the shell shape to the shape of the boundary that confines it.  We expect these results to be relevant for an expanse of shell shapes, and thus apply to the design of shape--shifting materials and the swelling and growth of soft structures.
\end{abstract}
\pgfplotsset{compat=newest}

\maketitle

\begin{figure}
	\centering
	\includegraphics[width=.6\linewidth]{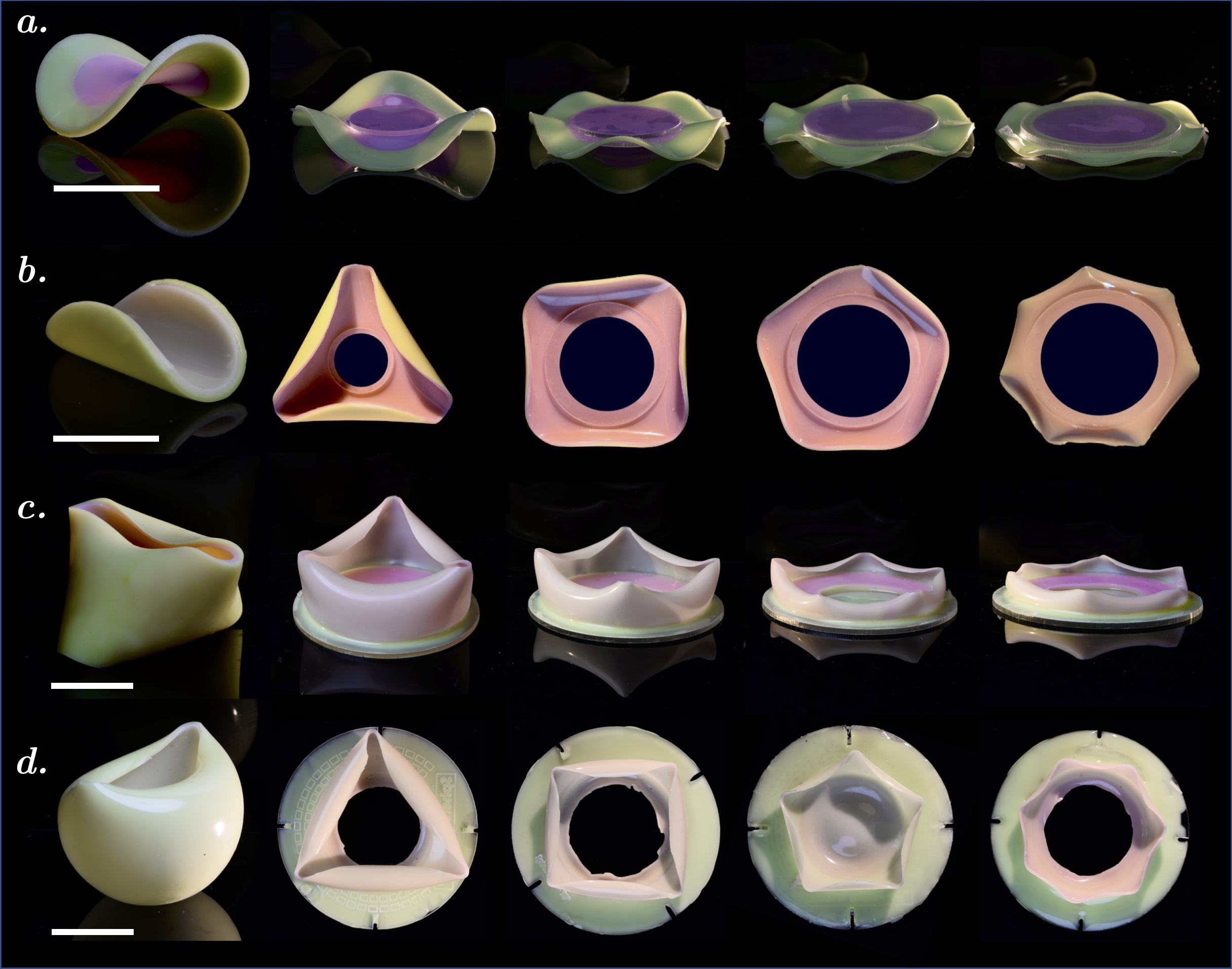}
	\caption{As the extent of confinement increases from left to right, the $a.$ saddle, $b.$ rolled sheet, $c.$ cylinder and $d.$ spherical segment exhibit more lobes. In $a.$ and $b.$, shells are clamped between acrylic plates of increasing diameter. The cylinder and sphere cut to varied $c.$ heights and $d.$ latitudes, and fixed with an acrylic ring at the base. Shells are made of polyvinylsiloxane (PVS). Scale bars represent 30 mm.}
	\label{fig1}
\end{figure}

Shells are notorious for their nonlinear response to mechanical loading, and subtle changes to how they are held, or constrained, can have profound affects on how they deform. Confinement of soft shells can induce dramatic deformations as illustrated in Fig.~\ref{fig1}, where radial confinement is increased from left to right. These mechanics are relevant to soft biological tissues, as their morphology often depends on a combination of mechanical forces imparted along their boundaries, and non--mechanical forces that drive growth or swelling.  Confinement of soft tissues can result in the wrinkling and scar formation of surgical wounds~\cite{cerda2005}, and these changes in shape or morphology are not purely cosmetic. For example, during the embryogenesis of the ciliary body of an avian eye, differential growth induces wrinkles that radiate outward from the retina~\cite{bard1992}, a stiff region that resists deformation. Capillary blood vessels form in the valleys of these wrinkles, while molecules that promote neural cell adhesion fail to express in the regions where these epithelial tissues wrinkle~\cite{bard1982a, bard1982b}. These effects are entirely mechanical, as evidenced by experiments that induced wrinkles in the chick eyes by swelling them in ethanol~\cite{bard1992, bard1982b}. Similar studies on the differential swelling and growth of artificial tumors~\cite{dervaux11, dervaux2011b} and biofilms\cite{benamar14, dervaux14} described the role of confinement and the mechanics of these circumferential wrinkles in much greater detail. Radial confinement occurring within airways and arteries~\cite{goriely17}, as seen in buckling and folding of mucous membranes, can cause the collapse or closure of the oesophagus~\cite{yang2007}, blood vessels~\cite{lee1979}, and gastrointestinal tract~\cite{lu2005}.

\begin{figure}[h!]
	\centering
	\includegraphics[width=.7\linewidth]{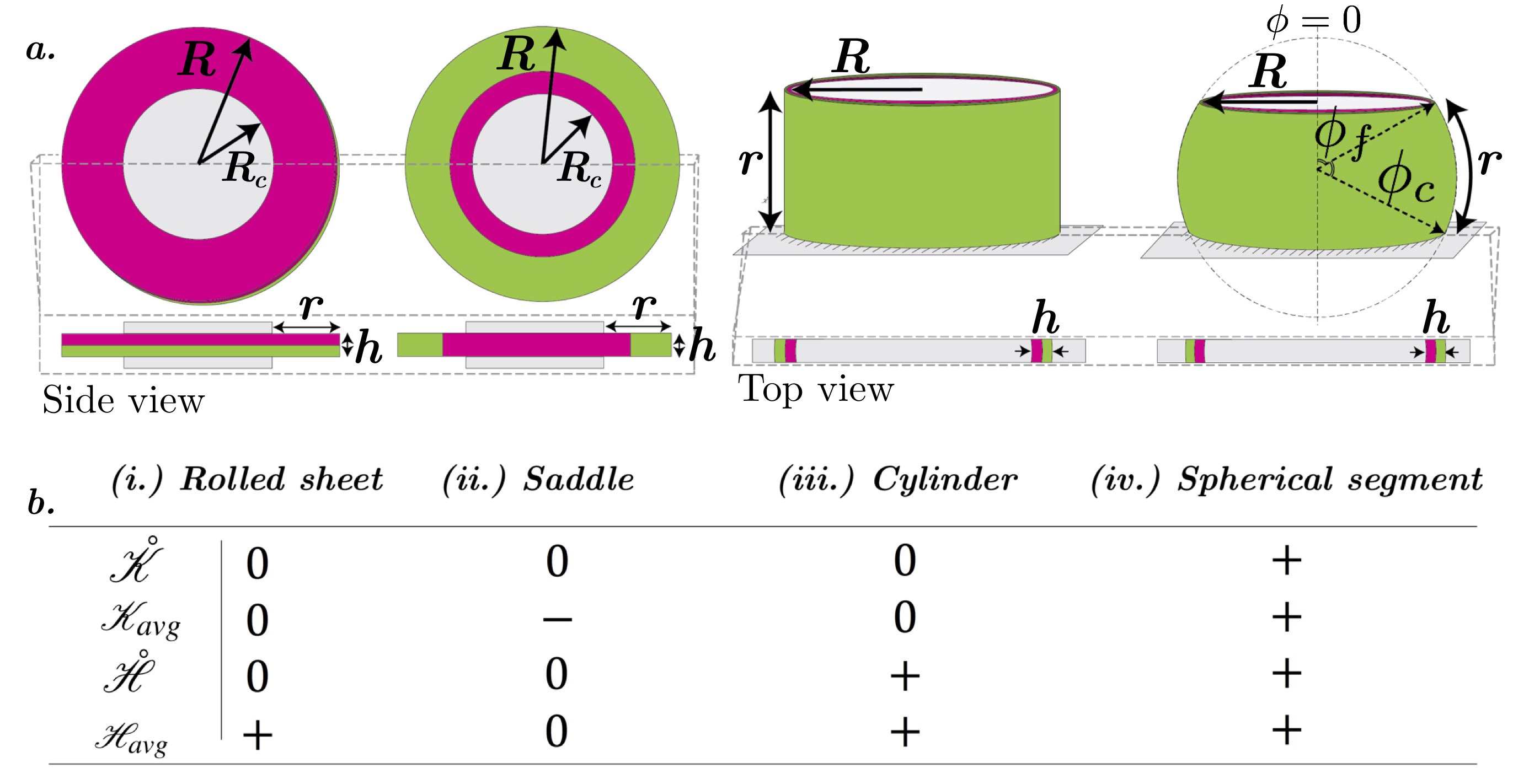}
	\caption{$a.$ Schematics showing pre-residual swelling configurations and relevant geometric parameters for the $i.$ rolled sheet, $ii.$ saddle, $iii.$ cylinder, and $iv.$ spherical segment. Pink areas will ``shrink" while green ones will "grow" upon residual swelling, and grey represents areas constrained by acrylic plates. $b.$ Table displays initial (accented "o") and post-swelling (subscripted "$avg$") Gaussian $\mathcal{K}$ and mean $\mathcal{H}$ curvatures for each shape.}
	\label{fig2}
\end{figure}

Beyond these biological systems, the ability to prescribe and control the shape of objects has ushered in an age of {\em designer materials}~\cite{Reis2015}. By dictating the volumetric strain in specific regions of soft elastomers, researchers have been able to morph 2D sheets into 3D shells~\cite{Klein2007b, Holmes2011, Kim2012a}, with features spanning multiple length scales~\cite{Dervaux2012, Pandey2013}. Differential swelling, sometimes accomplished by using the residual polymer chains left in portions of cured elastomers, has been used to fabricate helical ribbons~\cite{Wu2013}, rolled sheets~\cite{pezzulla16}, saddles~\cite{pezzulla15}, pinched spheres~\cite{pezzulla18}, and wavy strips and discs~\cite{Mora2006, DuPont2010, Barros2012}. Even for free, unconstrained plates and shells the shape selection process is non--trivial. The shapes that result from differential swelling can be determined by examining how swelling alters the metric tensor of the middle surface of the plate, an approach described by the so--called theory of unconstrained non--Euclidean plates~\cite{efrati2009}. When swelling only imparts a local curvature change along the middle surface, as is the case for the residual swelling of bilayer plates and shells, the non--mechanical swelling process can be cast as a mechanical stimulus which alters the natural curvature of the shell~\cite{pezzulla18}, and the stability of these structures can be evaluated using techniques common to applied mechanics. The inverse problem -- knowing a desired shape and searching for the correct initial conditions necessary to achieve it --  is a problem that has received far less attention, but will likely be more desirable. Work by Dias {\em et al.} demonstrated how to find the metric for a variety of axisymmetric shapes~\cite{dias2011}, while more recent work has shown how to find the metric for a wide range of shapes, including a human face, when a negative curvature can be prescribed at any point~\cite{vanRees2017}.

In the effort to understand and control shape change in soft and thin structures, the interplay between intrinsic geometry and geometric constraints is still not well understood. Confining a simple 1D object, {\em i.e.} an {\em elastica}, within a rectangular box is a nontrivial problem, in part due to the unknown and evolving location of the point of contact between the elastica and the walls~\cite{domokos97, holmes99, roman99, roman02, pocheau04}. Similar problems emerge in the packing of thin sheets, for instance pushing a plate through a ring causes it to form a {\em developable} cone, or {\em d}--cone~\cite{BenAmar1997, Cerda1998, Chaieb1998, Cerda1999, Chaieb2000, cerda05}, and in the confinement of a thin plate between two hemispheres~\cite{hure12} or onto a droplet of water~\cite{Paulsen2015}. Confinement of intrinsically curved shells has received less attention, with an exception being the behavior of shells under indentation~\cite{vaziri08}, including a hybrid experimental-numerical study of the response of positively curved shells to indenters of varied geometries~\cite{nasto13}. In this work, we present a primarily experimental study on how geometric confinement facilitates pattern formation in structures with intrinsic curvature. We consider shells with various mean and Gaussian curvatures under a range of radial confinement, and we examine the combined role of radial and transverse confinement on negatively curved shells, or saddles.  We focus our study on four categories of shapes, shown in Fig. \ref{fig1} and schematically in Fig. \ref{fig2}a: saddles, rolled sheets, cylinders, and spherical segments. Each is initially axisymmetric and exhibits periodic postbuckling patterns when subjected to geometric constraints. These geometries were chosen to access a range of average Gaussian and mean curvatures in their reference ($\mathring{\mathcal{K}}_{avg}$, $\mathring{\mathcal{H}}_{avg}$) and deformed ($\mathcal{K}_{avg}$, $\mathcal{H}_{avg}$) configurations.

The range of structures studied is outlined in Fig. \ref{fig2}b. Of the four shapes we study, two start out as flat plates -- one of these changes its average mean curvature, and the other changes its average Gaussian curvature after fabrication. Our study omits spherical caps, which have been well-characterized, {\em e.g.} in Ref.\cite{nasto13}. 

These shells are relatively thick as compared to recent work on thin film confinement~\cite{Paulsen2015, Vella2018, Paulsen2018}, and we will show that the characteristic pattern of deformation can be described by a single geometric parameter that appears to be independent of shell thickness in this regime.

\section{Radial Confinement} \label{radcon}

\begin{figure}[h]
	\centering
	\includegraphics[width=.5\linewidth]{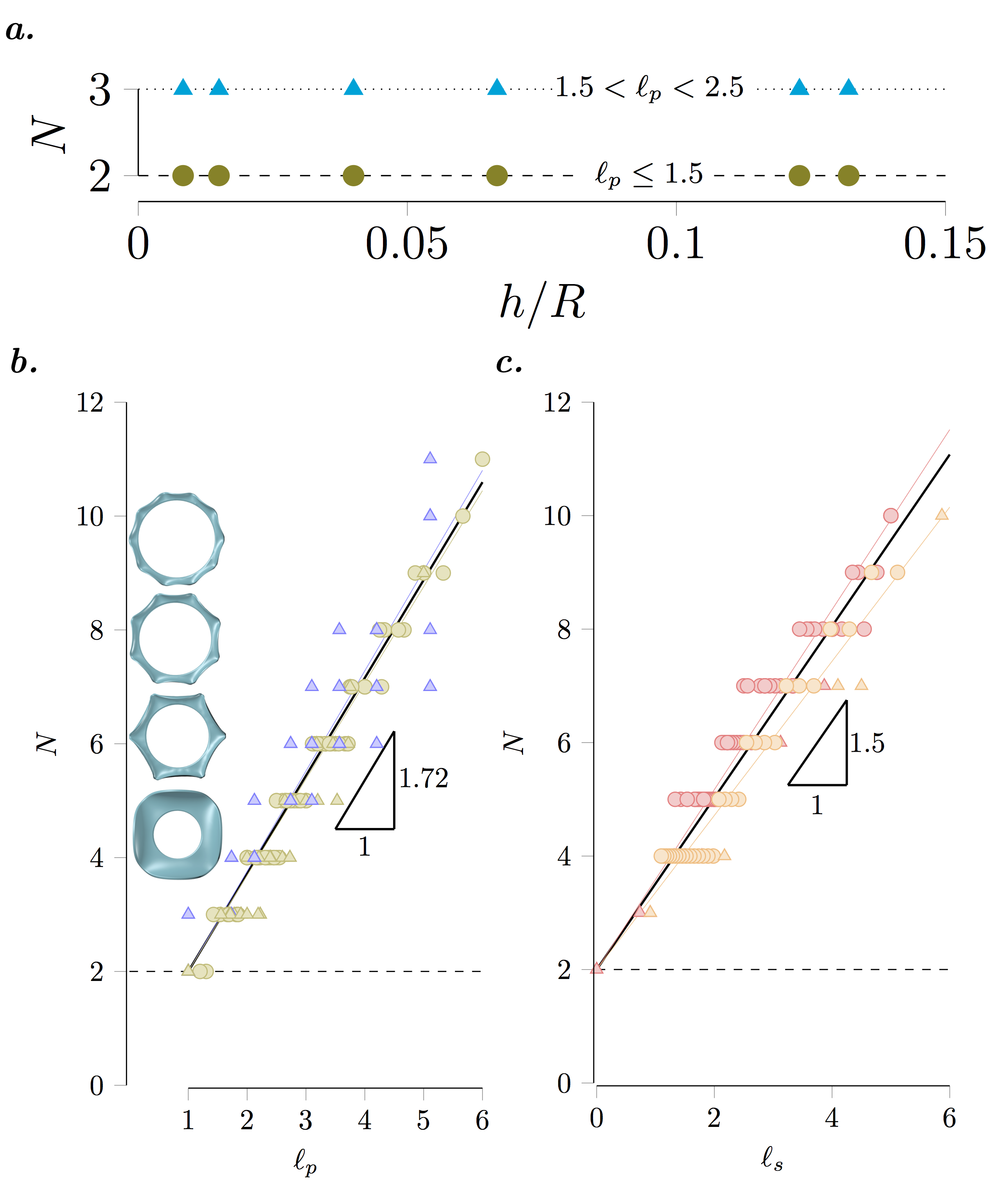}
	\caption{$a.$ In the thickness range we study, the wavenumber is insensitive to changes in thickness, $h$. Instead, the amount of unconstrained material, quantified by $\ell$, sets the wavenumber: for $\ell_p$ in a fixed range, but $h/R$ varied, $N$ (shown for rolled sheets) is unaffected. $b.$\&$c.$ The number of lobes $N$ may be reduced to one geometric parameter, $\ell = R/r$, which quantifies the relative amount of constraint. The evolution of $N$ is a linear function of $\ell$. Triangles are experimental data points and circles are from simulations. Solid lines of best fit, and their slopes, are shown in each plot. Free of constraints, each shape has two lobes (dotted line). Solid horizontal axes are drawn according to the minimum value for $\ell$. $b.$ For saddles (purple) and rolled sheets (green), $\ell_p = R/(R-R_c)$, and $min(\ell_p)=1$. Inset: Results from simulations for rolled sheets. $c$. For cylindrical shells (red) with height $r$, $\ell_s = R/r$. For spherical segments (orange), $\ell_s=\sin \phi_f/(\phi_c-\phi_f)$. The minimum of $\ell_s$ is $0$.}
	\label{scaling}
\end{figure}

Each of the unconstrained, residually stressed shapes are shown in the leftmost column of Fig.~\ref{fig1}. We begin by constraining the rolled sheets and saddles in the radial direction by clamping the shells between two rigid acrylic sheets of radius $R_c$. In Fig.~\ref{fig1}a\&b, we increase $R_c$ from left to right while keeping the shell radius $R$ and thickness $h$ constant, and we see that the number of lobes $N$, or wavenumber, increases. This wavenumber appears to be insensitive to changes in thickness in the range of $h/R$ we considered (Fig.~\ref{scaling}a), {\em $h/R \in [0.008, 0.13]$} -- thicker shells behave more like 3D bodies, while thinner shells made from these materials (see: Supplemental Information) deform significantly under gravity. Instead, it appears that the wavenumber is inversely proportional to the length of material that is unconstrained, {\em i.e.} $r \equiv R-R_c$. 
It is typical in wrinkling problems to expect the wrinkle wavelength $\lambda \sim N^{-1}$ to be proportional to a balance of the bending rigidity $B$ to the substrate, or foundation rigidity $K$~\cite{cerda05}, with $\lambda \sim (B/K)^{1/4}$. The bending rigidity appears in the bending energy of the shell, which we assume to be decoupled from the stretching energy, and is known to scale as $\mathcal{U}_b \sim \frac{B}{2} \int \kappa^2 \text{d}\omega$, where $\kappa$ is curvature, and $\text{d}\omega$ is the area element. The bending energy penalizes high curvatures, so from the perspective of the elastic energy, long wavelengths are preferable. The question in regard to these constrained shells is: what contributes to the foundation rigidity? We know that the strain energy of the foundation resisting deformation consists of a foundation stiffness $K$ that penalizes large amplitudes of deformation $A$, meaning that short wavelengths are preferable. Since in this thickness regime the shell thickness does not appear to play a dominant role in setting the wavelength, or wavenumber, of these constrained shells, we expect that any balance between bending rigidity and foundation rigidity should be independent of thickness to leading order. We hypothesize that the free length of the shell can be considered a series of cantilevers of length $L$, clamped at the the constraint, and bending with an unknown deflection $A$ at the free end. In this configuration, the cantilevers that are ``cut'' from the constrained shell have a bending stiffness of $K\sim B/r^3$. In this configuration, the strain energy of the foundation is given by:
\begin{equation}\label{Uk}
\mathcal{U}_K \sim K \int A^2 \ \text{d}L \sim \frac{B}{r^3} A^2 \cdot r.
\end{equation}
We assume that the wrinkle curvature will scale with the amplitude and wavelength as $\kappa \sim \frac{A}{\lambda^2}$, and that for simplicity the unconstrained area of the shell will scale as $\omega \sim r^2$, such that 
\begin{equation}\label{Ub}
\mathcal{U}_b \sim \frac{B}{2} \frac{A^2}{\lambda^4} \cdot r^2.
\end{equation}
Balancing the two energies in \eqref{Ub} and \eqref{Uk} gives $\lambda \sim r$. With $\lambda = 2 \pi R/N\sim R/N$, and defining $\ell \equiv R/r$, we arrive at a scaling of the wavenumber as a function of the unconstrained, or free length of the shell
\begin{equation}\label{eqscaling}
N \sim \ell.
\end{equation}
In Fig.~\ref{scaling}b, we plot experimentally and numerically obtained wavenumbers $N$ as a function $\ell_p$, which is $\ell$ for the shells that initially started as flat plates (see Supplemental Information for details on fabrication and simulations). When the constraint $R_c \rightarrow 0$ the dimensionless length $\ell_p \rightarrow 1$, and experiments on unconstrained shells confirm that $N \rightarrow 2$ (Fig.~\ref{fig1}), suggesting that for rolled sheets and saddles equation~\ref{eqscaling} should be modified to $N\sim\ell_p+1$. This scaling is plotted as a solid line on Fig.~\ref{scaling}b, with a slope of 1.72 found via linear regression. We would expect the slope to be of $\mathcal{O}(1)$ if the scaling is valid, and these results suggest that our approximation of the foundation energy was reasonable. 

We now turn our attention to shells with initially nonzero mean and Gaussian curvatures. Physically, the scaling from equation~\ref{eqscaling} suggests that the wavenumber will increase linearly as the free, unconstrained length of the shell decreases. For the cylinders and spherical segments constrained at their base, the free length that decreases from left to right on Fig.~\ref{fig1}c\&d is the arclength $r$ of material from the clamped base to the shell opening, and similar to the rolled sheets and saddles, as this free length is decreased the wavenumber increases. Therefore, we anticipate that equation~\ref{eqscaling} will also capture the wavenumber of these constrained shells, provided the appropriate limits on $N$ and $\ell_s$ are met, where $\ell_s$ is $\ell$ for shells that have are initially curved. Here, the unconstrained shell corresponds to $r\rightarrow \infty$, or $\ell_s \rightarrow 0$, which is analogous to the two lobe deformation ({\em i.e.} $N \rightarrow 2$) observed with a ``pinch in a pipe''~\cite{mahadevan07}. This suggests that for cylinders and spherical segments, we expect that equation~\ref{eqscaling} should be modified to $N\sim\ell_s+2$. In Fig.~\ref{scaling}c, we plot experimentally and numerically obtained wavenumbers $N$ as a function $\ell_s$ for cylinders and spherical segments. The scaling $N\sim\ell_s+2$ is plotted on Fig.~\ref{scaling}c, with a slope of 1.50 found via linear regression. These results seem to be in good agreement with this reduced order model, suggesting that the wavenumber of a wide range of constrained shells can be characterized with a dimensionless parameter corresponding to the free length of the shell.

\section{Transverse confinement} \label{transcon}
\begin{figure}[H]
	\centering
	\includegraphics[width=.5\linewidth]{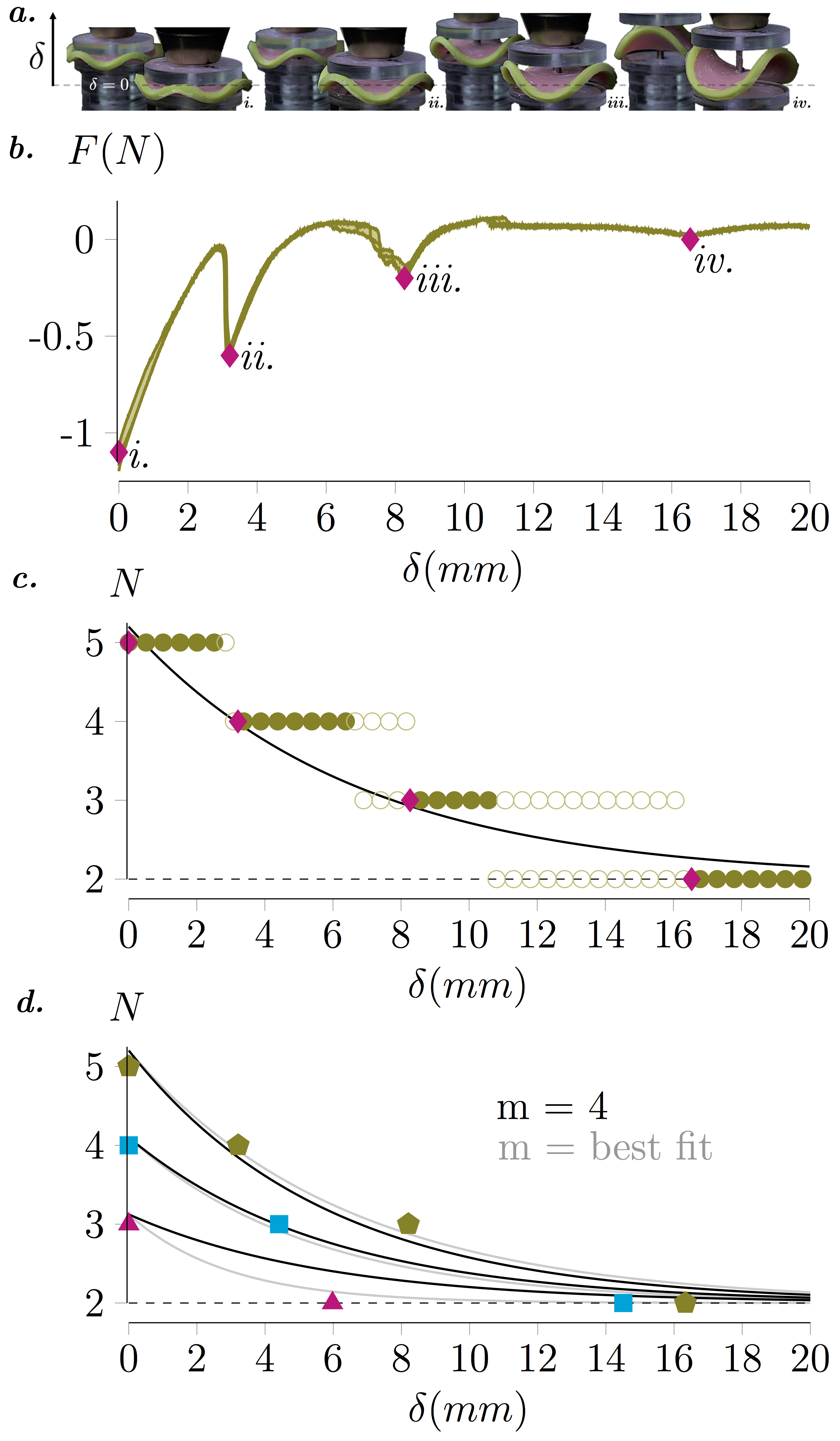}
	\caption{$a.$ As $\delta$ increases from left to right, the number of lobes ($N$) decreases.  The recessed images are mirrored views showing the back side of saddles. $b.$ Force-displacement curves of three displacement-controlled tests on a single shell. Pink diamonds correspond to lobe transition points with Roman numerals indicating the transition from five (i.), to four (ii.), to three (iii.) to two (iv.) lobes. $c.$ Wavenumber $N$ vs. $\delta$ for the same sample. Solid points correspond to symmetric lobes, and open circles correspond to transitions between lobes. We expect the solid theoretical curve to capture the points marked by pink diamonds. This curve is equation \eqref{genscaling} with $m=3.5$. $d.$ The same equation captures experimental lobe switches (points, as in $c$) with different geometric parameters. Black curves fix $m=4$, and grey curves correspond to best fit $m$ values: from top, $m = 3.68$, $m=4.38$, $m=8.00$.
		\label{onetest}}
\end{figure}

We will now relax the radial confinement to investigate shell behavior under varying amounts of transverse confinement. We focus primarily on an experimental analysis of saddles, because to our knowledge there are only limited examples of the transverse confinement of saddles in the literature, and the experiments on saddles are the most practically feasible out of the structures discussed in section~\ref{radcon}. We constrain the shells in the transverse direction with quasi-static, displacement-controlled tests in which a saddle is compressed between pairs of acrylic plates of radius $R_c$. Initially, the distance $d$ between the top and bottom plates equals the saddle's thickness, {\em i.e.} $\delta=d-h$ (Fig.\ref{onetest}a\&b,i.), which represents the limit discussed in Section~\ref{radcon}.

As we separate the distance between the two plates by an amount $\delta$, there is a non-monotonic decrease in the applied compressive load, and the number of lobes decreases, as shown in Fig.~\ref{onetest}a, (a more detailed experimental protocol is discussed in the Supplemental Information). The decrease in the compressive load is nearly linear for $\delta/A << 1$, and then reaches a minimum when one point of contact between the acrylic plate and the saddle is lost, thus beginning the mode shape transition from $N=N_{\max}$ to $N-1$ lobes. The load immediately increases, and once the saddle has reaches $N-1$ (asymmetric) lobes (Fig.\ref{onetest}a\&b,ii.) the load once again decreases. When a new symmetric shape is reached at $N-1$ lobes, the slope of the force-displacement curve decreases but remains positive, and the process repeats until $\delta \simeq A$, $F \simeq 0$, and there are $N=2$ lobes (Fig.\ref{onetest}a\&b,iv.). The slope of the force-displacement curves through these transitions appears to gradually decrease as $N$ decreases. Fig.~\ref{onetest}b shows these trends in a force-displacement curve for a representative sample that achieves a maximum of $5$ lobes at $\delta=0$.

We now aim to provide some mechanistic insight as to the lobe transitions from $N_{\max}$ to $N=2$ for saddles as the transverse constraint is reduced. Here, we know the two limits: (1.) as $\delta \rightarrow \infty$ we expect that $N\rightarrow2$, and (2.) as $\delta \rightarrow 0$ we expect that $N\rightarrow N_{\max}$ as given by equation~\ref{eqscaling}. The first limit can be simplified, because the sheet will be unconstrained once $\delta$ is larger than the amplitude of the shell's lobes, {\em i.e.}  $N\rightarrow2$ when $\delta \geq A$. In these experiments, while in principle $\ell_p$ is fixed, in effect the free length of the shell may be approximated as being a function of $\delta$, {\em i.e.} $r_\delta(\delta)$, with $r_\delta(0) = r$ from section~\ref{radcon}. As an {\em ansatz} for $r_\delta$ that meets these two limits, we chose a logistic function in the form
\begin{equation}\label{fform}
r_\delta(\delta) = \frac{R}{1+\frac{R_c}{r} e^{-m\frac{\delta}{A}}},
\end{equation}
where $m$ is an unknown constant that describes how quickly the unconstrained length will transition between $r$ and $R$. Substituting this {\em ansatz} into equations $\mathcal{U}_b$ and $\mathcal{U}_k$ for $r$, we can solve for $N(\delta)$. Following some algebra, we find
\begin{equation}\label{genscaling}
N(\delta) \sim \ell_p + 1 - \frac{R_c}{r} \big(1- e^{-m\frac{\delta}{A}}\big). 
\end{equation}
which reduces to equation~\ref{eqscaling} for plates when $\delta=0$. In Fig.~\ref{onetest}c, we plot the experimentally observed wavenumber as a function of $\delta$. The transition process is highly nonlinear, and so we note the transition between two wavenumbers with open symbols, while highlighting the transition points from the local minima in the force-displacement curve as filled diamonds. Equation~\ref{genscaling} is plotted as a solid black curve, with $m=3.5$ chosen as a best fit parameter to the transition points. Although $m$ is effectively a fitting parameter, we anticipate that it will depend on the bending rigidity of the saddle. We have not taken into account how the magnitude of the saddle's Gaussian curvature, which will be related to the amplitude of the lobes, nor the shell thickness affect the transition points, however we expect that $m$ will be a function of these parameters. Further testing, in particular numerics, will help explain the contribution of $\mathcal{K}$ and $h$ to the transition between mode numbers. Still, we note that for the samples we tests, using $m=4$ captures the transition points for shells with $N_{\max}=5$, $N_{\max}=4$, and $N_{\max}=3$ (Fig.~\ref{onetest}d). Choosing the best fit values of $m$ for each sample ($m=3.7$, $m=4.4$, and $m=8.0$) changes the critical $\delta$ for observing lobe transitions, but qualitatively provides similar values. In general, the form of \eqref{genscaling} clarifies the relative contributions of transverse ($\delta$) and radial ($R_c$) confinement. At low $\delta$ values, radial effects dominate. As the wave number depends on $\delta$ exponentially, however, the effects of reduced transverse confinement quickly take over with increasing $\delta$. 

\section{Conclusions}
In this work, we explored geometry's fundamental role in the periodic buckling patterns that emerge in confined shells. We studied shells covering a range of Gaussian and mean curvatures, accessible via residual swelling. We first saw that one simple geometric parameter, $\ell$, which relates the overall shell radius to the amount of unconfined material, predicts the number of wrinkles ($N$) a radially confined shell will adopt. Then for negatively curved saddles, we reduced the radial constraint by varying transverse confinement and measured the transition points between wavenumbers. We observed that decreasing the amount of confinement, whether in-the-plane or vertically, can be interpreted as a reduction in stiffness to the free region of the shell. In a low-stiffness regime, lower buckling modes are energetically preferable. This interpretation allowed us to generalize $\ell$ to include our range of transverse confinement. Thus, the model given by relation \eqref{genscaling} captures a wide range of bidirectional confinement. 

There is much to be done in terms of more rigorously understanding why thickness appears to be unimportant, and to put the cantilever analogy on firmer ground. A nice analog to our transverse confinement of saddles is the transverse confinement of an elastica~\cite{roman99, roman02}. In these works, solutions for the confined elastica~\cite{holmes99,domokos97} are extended to thin plates constrained progressively in the vertical direction. Our problem has subtle differences, notably that our shells are naturally curved, and our confining plates are smaller than the shell size. However, the transitions between buckling modes in our experiments are reminiscent of these studies, including qualitative features like planar contact, free-standing folds, and rolling~\cite{pocheau04}. These parallels suggest a way to pursue a more formal connection between the two problems. The shells studied in this work are residually stressed, and the magnitude of residual stress did not enter our mechanical model. It was recently shown that the magnitude of residual stress in shells will alter the critical point at which an instability occurs, {\em i.e.} the load required to buckle the structures, but that the instability remains qualitatively similar~\cite{jiang18}. Also, as others~\cite{nasto13} have observed, contact plays an important role in transverse confinement. Further numerical analysis of these constrained shells would be beneficial, in particular, such an analyses could also offer a more geometric freedom, with regards to both shells and their confining boundaries, beyond what is readily accessible experimentally. In general, we anticipate these results will aid in the design of shape--shifting structures, and we believe there is much work to be done on understanding the role of confinement when designing structures that change shape on command.

\section*{Conflicts of interest}
There are no conflicts to declare.

\section*{Acknowledgements}
LSM and DPH gratefully acknowledge the financial support from NSF through CMMI-1824882.

\section{Supplementary Information}

\subsection{Structure Fabrication}

To fabricate the shapes shown in Fig.~\ref{fig1}, we use a technique known as residual swelling~\cite{pezzulla15, pezzulla16}. We use two polyvinylsiloxane (PVS) elastomers, which we will refer to as {\em green} (Zhermack Elite Double 32, E=0.96 MPa) and {\em pink} (Zhermack Elite Double 8, E=0.23 MPa). The materials are cast in as fluids and allowed to thermally crosslink, or cure, at room temperature for 20 minutes. After curing, the pink elastomer has residual polymer chains within the material, and these residual free chains flow into the green elastomer when the two materials are in contact with each other. The local loss of mass causes the pink material to decrease in volume, or shrink, while the green material correspondingly swells, thus inducing a \textit{differential} swelling in the structure which preserves its total mass. Differential swelling in shells can lead to residually stressed structures that emerge because the shell must deform to accommodate a geometric incompatibility~\cite{klein07}. When the differential swelling occurs through the shell's thickness, it deforms in a nearly isometric manner in the bulk of the shell, away from shell's edges~\cite{pezzulla18}, and when the differential swelling occurs in-the-plane of the shell the deformation is dominated by stretching~\cite{pezzulla15}. These opposing deformations explain why the initially flat shapes can be morphed into either rolled sheets or saddles. As residual swelling is a diffusive process, the time to deform scales with the square of the dimension across which swelling occurs. This characteristic dimension for swelling is either the thickness $h \approx 1mm$, or in the case of saddles, where residual swelling occurs in-the-plane, the radius, $R=30mm$~\cite{pezzulla15}.

To make homogeneous rolled sheets, we use a spin coater (Laurel Technologies, WS-650-23) to deposit a pink layer of PVS atop a laser-cut (Epilog Laser Helix, 75W) circular acrylic plate, $R \in [25mm, 35mm]$. After it cures, we add a green layer in the same manner. The residual swelling first bends the sheet into a shallow spherical cap, and then ultimately buckles it into a rolled sheet -- a cylinder--like shape that is open along its directrix. The rolled sheet is nearly isometric away from its edges ({\em i.e.} $K_{avg} = 0$) and its non--zero mean curvature is linearly proportional to the natural curvature imposed by residual swelling~\cite{pezzulla16, pezzulla17, pezzulla18}.

Saddles are made by laser-cutting a negative circular mold ($R=30mm$) from clear cast acrylic sheets of thickness $h$: $0.794$mm$\rpm 0.119$mm (Inventables), $1.589$mm (tolerance $-0.584$mm to $+0.254$mm), $2.381$mm ($-0.034$mm to $+0.025$mm), or $3.175$mm ($-0.635$mm to $+0.381$ mm) (McMaster-Carr). This circular mold is glued atop a base acrylic plate, and a smaller circle, radius $R_c \in \{12.5mm, 28.25mm\}$, is centered and fixed to the base plate. We then pour green PVS to form a ring, filling the mold up to the acrylic sheet thickness. After the ring cures, the smaller circle is removed and the remainder is filled with pink PVS. After residual swelling, a saddle shape forms: $H_{avg} \approx 0$ and $\mathring{K}_{avg} < 0$ -- the value of the latter depends on the ratio of pink to green polymer\cite{pezzulla16}. In-plane swelling is quite a bit slower than through-thickness swelling, since the characteristic length scale changes from the thickness to the radius~\cite{pezzulla15}. The dynamics can be increased by extracting the free polymer chains in a solvent bath, {\em e.g. ethyl acetate}.

\begin{figure}[h]
	\centering
	\includegraphics[width = .6\linewidth]{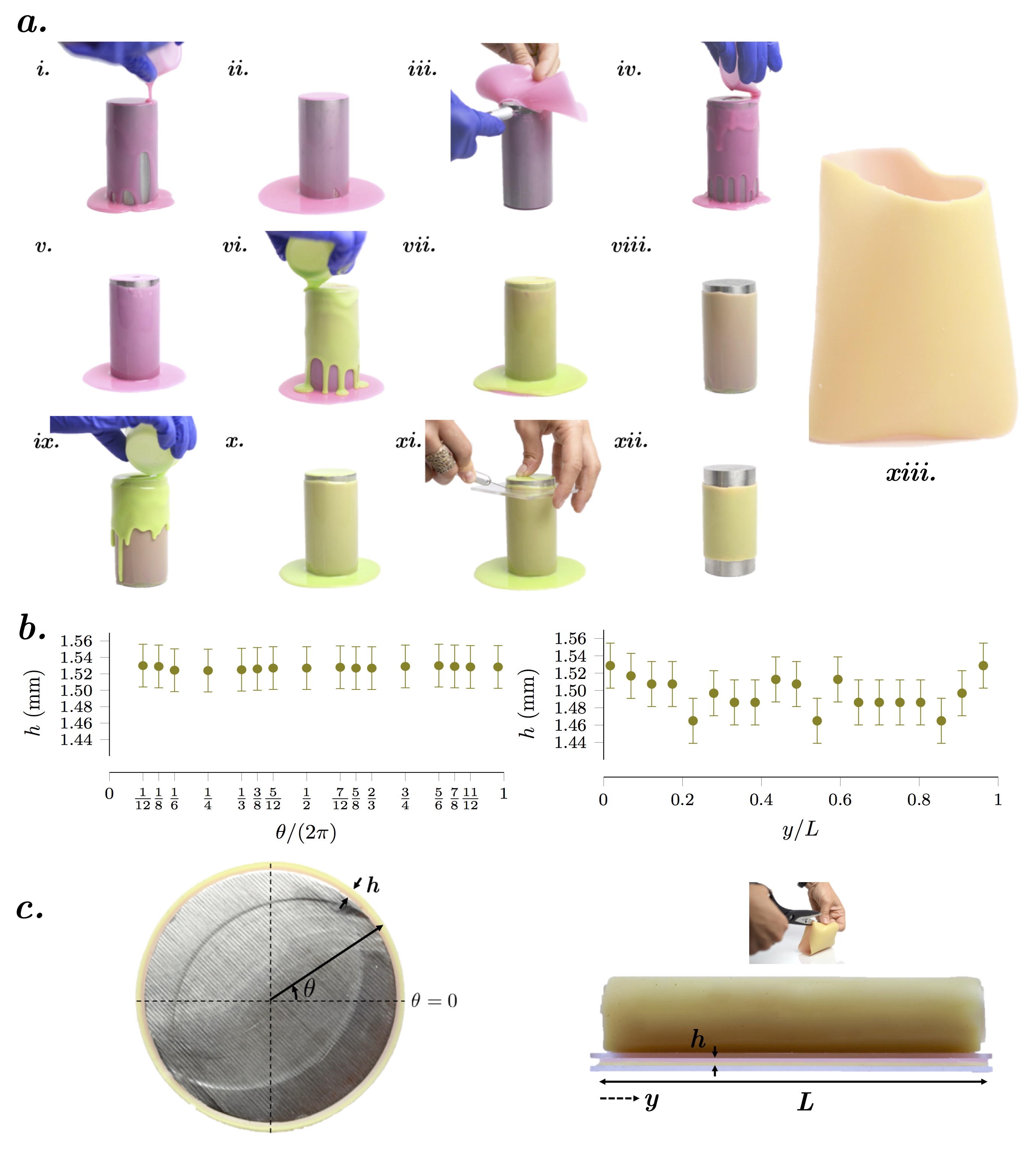}
	\caption{$a$. Fabrication process for a bilayer cylinder: (i.-ii.) The first pink layer of PVS is poured to uniformly coat the steel cylinder. (iii.) After 20 minutes, the first layer has cured. The cylinder is flipped upside-down, and the excess material is removed. (iv.-v.) A second layer of pink PVS is poured in order to achieve a uniform thickness in the vertical direction. (vi.-vii.) After another 20 minutes, both pink layers have cured. A green layer is added in the same manner. (viii.) After 20 minutes, the excess material is removed and the cylinder is flipped upside-down again. (ix.-x.) The final green layer is added. (xi.-xii.) After the final layer has cured (20 minutes), the bilayer cylinder is cut from the mold using a straightedge. (xiii.) The bilayer cylinder once peeled from the mold. Residual swelling causes the cylinder to buckle into a ``pinched pipe''. $b$. Thickness versus (right) axial and (left) azimuthal ($y=0$) position, corresponding to parameters shown in the images in $c$. Thickness is relatively uniform in both directions, albeit more so in the radial direction. Measurements were taken using ImageJ. $c$. Top view (left) and cut view (right) show parameters relevant to the above plots. Right: to obtain thickness measurements in the axial direction, the cylinder is cut, then glued between glass plates to prevent rolling.
	}
	\label{cylfab}
\end{figure}

Cylinders and spherical segments are poured as bilayers over corresponding 3D molds. Spherical segments are formed by coating a metal ball-bearing with viscous PVS so that each layer has approximately uniform thickness~\cite{lee16, pezzulla18}. These spherical shells have positive average mean and Gaussian curvatures both before and after the swelling process. Cylinders are fabricated similarly~\cite{marthelot17} (see Fig.\ref{cylfab}), and like spheres, the initial mean curvature $\mathring{\mathcal{H}} > 0$, though $\mathring{\mathcal{K}} = 0$. After deformation, a "pinched pipe" forms \cite{mahadevan07}, with $\mathcal{H}_{avg}>0$ and $\mathcal{K}_{avg} \approx 0$.

For the experiments described in Section \ref{radcon}, rolled sheets and saddles are clamped in the center between two laser-cut acrylic plates of equal size, $R_c \in \{12.7mm, 30.5mm\}$. Cylinders and spherical segments, on the other hand, are constrained by acrylic plates glued to the base with a very thin layer of green VPS. Cylinders are then cut to varied heights, and spheres are cut at different latitudes: the angle $\phi_f$ is subtended by the arclength from the origin (the north pole) to the top cut (the free surface). The base, where the shell is constrained, is defined by the angle $\phi_c$. Schematics of the pre-swelling configurations, including constraints, are given in Fig. \ref{fig2}. Thickness is measured at $ h = \{0.25, 0.45,0.75,1.5,4\}$mm $\rpm 0.15$mm for Section \ref{radcon}, and $h \in [2.381$mm $\rpm 0.1$mm, $3.175$mm $\rpm 0.1$mm] for the saddles discussed in Section \ref{transcon}.

\subsection{Mechanical force testing}
The saddles used in Section \ref{transcon} were fabricated with a centered $2.25$mm radius hole through which we guide a $2$mm radius rod as transverse confinement is varied. We determined this hole to be necessary for maintaining the saddle's position but negligible for our purposes -- it has no effect on lobe number. 

We investigate transverse confinement with a setup designed for the INSTRON 5943. We attach a drill-type grip (Instron 0.375in Keyless Drill-Type Chuck Assembly) to the load cell to secure an aluminum rod ($2$mm radius), which is screwed to an internally threaded acrylic plate of radius $R_c$. A second partially threaded rod is attached to the underside of this top plate, pointing downward. The rod is guided through the saddle's center hole and then through a hole also of radius $2.25$mm in the center of the acrylic base plate, which itself has radius $R_c$. The base plate is affixed to a thick tube of outer radius $<R_c$, inner radius $3.5$mm and height $44.45$mm. This tube is comprised of stacked acrylic rings each of thickness $6.35$mm, glued together and closed at the base. The base of the tube is screwed to a tapped optical table. 

Displacement-controlled tension tests are performed at a rate of 4 mm/min and force is measured with a 500N load cell (resolution 0.0025N). Videos were taken with a Nikon D610 DSLR Camera and were used for post-processing in conjunction with Instron data.

\subsection{Numerics}
For the three bilayer geometries where residual swelling occurs through-the-thickness, we sought to validate the experiments from Section \ref{radcon} with simulations developed in COMSOL Multiphysics. We created a 3D model within the context of three-dimensional neo-Hookean elasticity with large distortions using a Neo-Hookean material model~\cite{lucantonio14}. The stimulus is represented by a spherical distortion field $\vec{F}_\textup{o}=\alpha(\eta^3)\vec{I}$, with $\alpha(\eta^3)=\lambda_o$ in one layer, and $\alpha(\eta^3)=2-\lambda_o$ in the other layer, where $\eta^3$ is the general coordinate normal to the midsurface (i.e. across the thickness) and $\lambda_o$ is an inelastic stretching factor. The constraints were modeled with Dirichlet boundary conditions imposed around a ring in each case, reflecting the experimental setup.


\subsection{Buckling dynamics} \label{bucklingdynamics}
We know from Ref.~\cite{pezzulla17} that for the unconstrained rolled sheet ($\ell_p = 1$), the critical buckling curvature (normalized by the thickness) is

\begin{equation}\label{MPkappac}
\kappa_{b}h=\sqrt{10+7\sqrt{2}} \Big(\frac{h}{R}\Big)^2.
\end{equation}

\begin{figure}[h]
	\centering
	\includegraphics[width=.5\linewidth]{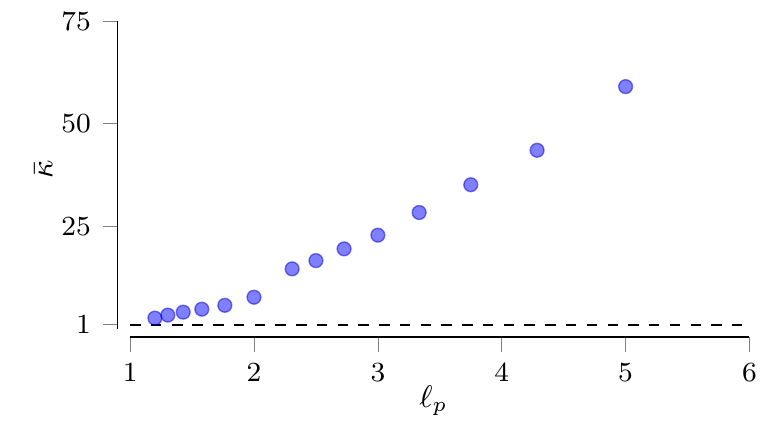}
	\caption{The critical buckling curvature for rolled sheets increases with the extent of constraint, quantified by $\ell_p$. $\bar{\kappa}$ is the ratio of the critical buckling curvature to that of the unconstrained case, from \eqref{MPkappac}.}
	\label{kappabuckling}
\end{figure}

We examine the effect of constraint on this value -- our numerics allow us to extract the critical curvature required for buckling for various values of $\ell_p$. This is shown in Fig. \ref{kappabuckling} by $\bar{\kappa}$, which represents the ratio of the buckling curvature of the constrained structure over an unconstrained but otherwise geometrically identical sheet (according to \eqref{MPkappac}). We observe that the buckling curvature increases with $\ell_p$ or, in other words, that more constrained structures require more curvature to buckle. 

\begin{figure}[h]
	\centering
	\includegraphics[width=.6\linewidth]{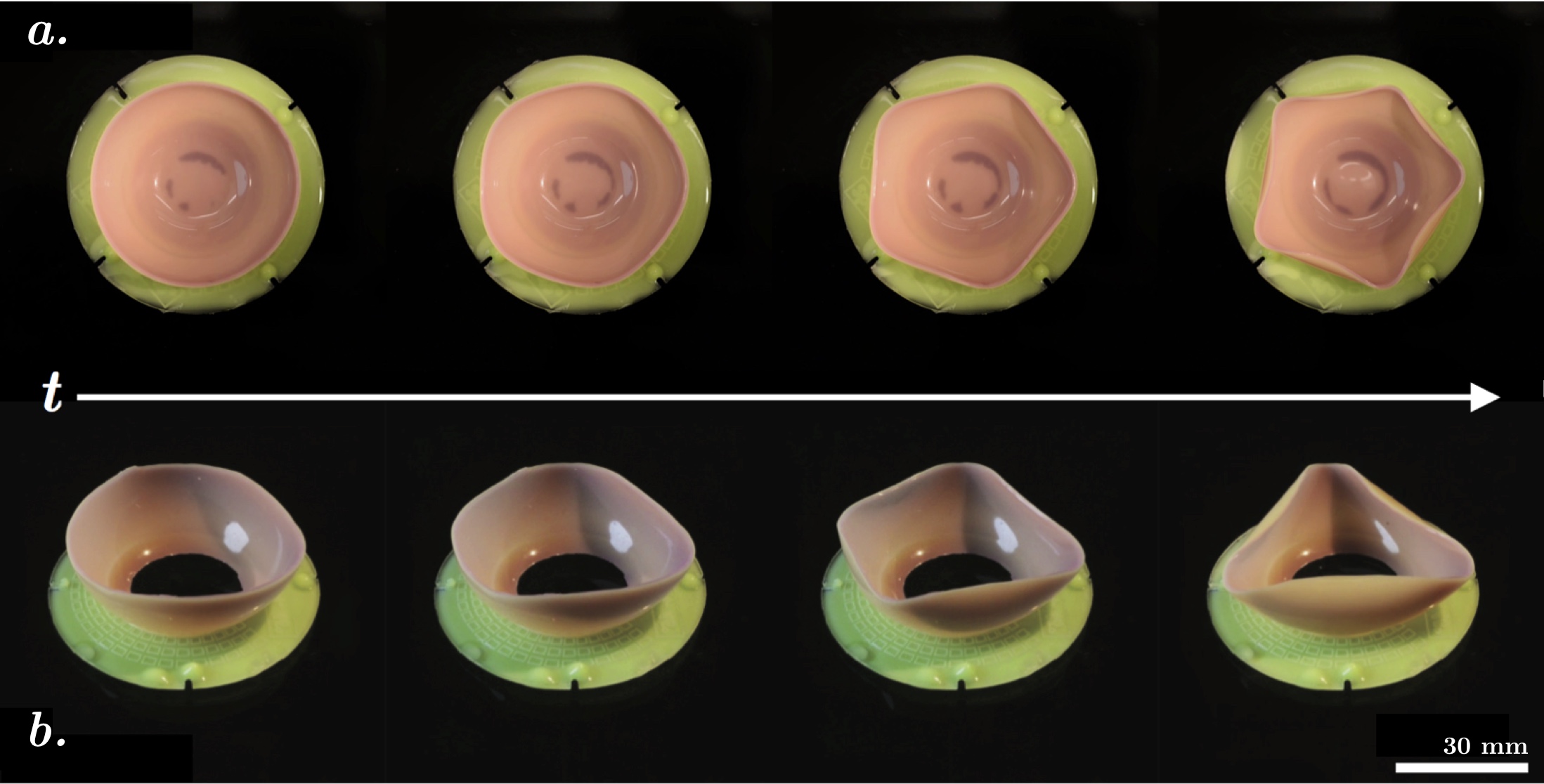}
	\caption{Timelapse images of constrained swelling spherical segments. Time between photos is about 40 minutes. $a.$ As curvature develops, a spherical segment ($\ell \approx 2$) develops five increasingly curved lobes. $b.$ A different sample ($\ell \approx 1$) is unstable through the residual swelling process: at lower curvatures (earlier in the residual swelling process), four lobes emerge. As curvature continues to increase, the spherical segment settles on three lobes as its energy-minimizing configuration.}
	\label{swellinglapse}
\end{figure}

Curvature continues to develop past the critical buckling point, and we generally observe that lobes become increasingly pronounced, as in Fig.~\ref{swellinglapse}a. However, when $\ell$ corresponds to an intermediate $N$ value, the lobe-selection process can be unstable (Fig.~\ref{swellinglapse}b.) A similar bistability between two mode numbers is seen in some shells after residual swelling is complete.

\begin{thebibliography}{52}
\providecommand*{\mcitethebibliography}{\thebibliography}
\csname @ifundefined\endcsname{endmcitethebibliography}
{\let\endmcitethebibliography\endthebibliography}{}
	\providecommand*{\natexlab}[1]{#1}
	\providecommand*{\mciteSetBstSublistMode}[1]{}
	\providecommand*{\mciteSetBstMaxWidthForm}[2]{}
	\providecommand*{\mciteBstWouldAddEndPuncttrue}
	{\def\EndOfBibitem{\unskip.}}
	\providecommand*{\mciteBstWouldAddEndPunctfalse}
	{\let\EndOfBibitem\relax}
	\providecommand*{\mciteSetBstMidEndSepPunct}[3]{}
	\providecommand*{\mciteSetBstSublistLabelBeginEnd}[3]{}
	\providecommand*{\EndOfBibitem}{}
	\mciteSetBstSublistMode{f}
	\mciteSetBstMaxWidthForm{subitem}
	{(\emph{\alph{mcitesubitemcount}})}
	\mciteSetBstSublistLabelBeginEnd{\mcitemaxwidthsubitemform\space}
	{\relax}{\relax}
	
	\bibitem[Cerda(2005)]{cerda2005}
	E.~Cerda, \emph{Journal of Biomechanics}, 2005, \textbf{38}, 1598--1603\relax
	\mciteBstWouldAddEndPuncttrue
	\mciteSetBstMidEndSepPunct{\mcitedefaultmidpunct}
	{\mcitedefaultendpunct}{\mcitedefaultseppunct}\relax
	\EndOfBibitem
	\bibitem[Bard and Bard(1992)]{bard1992}
	J.~B. Bard and J.~Bard, \emph{Morphogenesis: the cellular and molecular
		processes of developmental anatomy}, Cambridge University Press, 1992,
	vol.~23\relax
	\mciteBstWouldAddEndPuncttrue
	\mciteSetBstMidEndSepPunct{\mcitedefaultmidpunct}
	{\mcitedefaultendpunct}{\mcitedefaultseppunct}\relax
	\EndOfBibitem
	\bibitem[Bard and Ross(1982)]{bard1982a}
	J.~B. Bard and A.~S. Ross, \emph{Developmental Biology}, 1982, \textbf{92},
	73--86\relax
	\mciteBstWouldAddEndPuncttrue
	\mciteSetBstMidEndSepPunct{\mcitedefaultmidpunct}
	{\mcitedefaultendpunct}{\mcitedefaultseppunct}\relax
	\EndOfBibitem
	\bibitem[Bard and Ross(1982)]{bard1982b}
	J.~B. Bard and A.~S. Ross, \emph{Developmental Biology}, 1982, \textbf{92},
	87--96\relax
	\mciteBstWouldAddEndPuncttrue
	\mciteSetBstMidEndSepPunct{\mcitedefaultmidpunct}
	{\mcitedefaultendpunct}{\mcitedefaultseppunct}\relax
	\EndOfBibitem
	\bibitem[Dervaux and Ben~Amar(2011)]{dervaux11}
	J.~Dervaux and M.~Ben~Amar, \emph{Journal of the Mechanics and Physics of
		Solids}, 2011, \textbf{59}, 538--560\relax
	\mciteBstWouldAddEndPuncttrue
	\mciteSetBstMidEndSepPunct{\mcitedefaultmidpunct}
	{\mcitedefaultendpunct}{\mcitedefaultseppunct}\relax
	\EndOfBibitem
	\bibitem[Dervaux \emph{et~al.}(2011)Dervaux, Couder, Guedeau-Boudeville, and
	Amar]{dervaux2011b}
	J.~Dervaux, Y.~Couder, M.-A. Guedeau-Boudeville and M.~B. Amar, \emph{Physical
		Review Letters}, 2011, \textbf{107}, 018103\relax
	\mciteBstWouldAddEndPuncttrue
	\mciteSetBstMidEndSepPunct{\mcitedefaultmidpunct}
	{\mcitedefaultendpunct}{\mcitedefaultseppunct}\relax
	\EndOfBibitem
	\bibitem[Ben~Amar and Wu(2014)]{benamar14}
	M.~Ben~Amar and M.~Wu, \emph{EPL (Europhysics Letters)}, 2014, \textbf{108},
	38003--7\relax
	\mciteBstWouldAddEndPuncttrue
	\mciteSetBstMidEndSepPunct{\mcitedefaultmidpunct}
	{\mcitedefaultendpunct}{\mcitedefaultseppunct}\relax
	\EndOfBibitem
	\bibitem[Dervaux \emph{et~al.}(2014)Dervaux, Magniez, and Libchaber]{dervaux14}
	J.~Dervaux, J.~C. Magniez and A.~Libchaber, \emph{Interface Focus}, 2014,
	\textbf{4}, 20130051--20130051\relax
	\mciteBstWouldAddEndPuncttrue
	\mciteSetBstMidEndSepPunct{\mcitedefaultmidpunct}
	{\mcitedefaultendpunct}{\mcitedefaultseppunct}\relax
	\EndOfBibitem
	\bibitem[Goriely(2017)]{goriely17}
	A.~Goriely, \emph{{The Mathematics and Mechanics of Biological Growth}},
	Springer, 2017\relax
	\mciteBstWouldAddEndPuncttrue
	\mciteSetBstMidEndSepPunct{\mcitedefaultmidpunct}
	{\mcitedefaultendpunct}{\mcitedefaultseppunct}\relax
	\EndOfBibitem
	\bibitem[Yang \emph{et~al.}(2007)Yang, Fung, Chian, and Chong]{yang2007}
	W.~Yang, T.~Fung, K.~Chian and C.~Chong, \emph{Journal of Biomechanics}, 2007,
	\textbf{40}, 481--490\relax
	\mciteBstWouldAddEndPuncttrue
	\mciteSetBstMidEndSepPunct{\mcitedefaultmidpunct}
	{\mcitedefaultendpunct}{\mcitedefaultseppunct}\relax
	\EndOfBibitem
	\bibitem[Lee and Chien(1979)]{lee1979}
	M.~M. Lee and S.~Chien, \emph{The Anatomical Record}, 1979, \textbf{194},
	1--14\relax
	\mciteBstWouldAddEndPuncttrue
	\mciteSetBstMidEndSepPunct{\mcitedefaultmidpunct}
	{\mcitedefaultendpunct}{\mcitedefaultseppunct}\relax
	\EndOfBibitem
	\bibitem[Lu \emph{et~al.}(2005)Lu, Zhao, and Gregersen]{lu2005}
	X.~Lu, J.~Zhao and H.~Gregersen, \emph{Journal of Biomechanics}, 2005,
	\textbf{38}, 417--426\relax
	\mciteBstWouldAddEndPuncttrue
	\mciteSetBstMidEndSepPunct{\mcitedefaultmidpunct}
	{\mcitedefaultendpunct}{\mcitedefaultseppunct}\relax
	\EndOfBibitem
	\bibitem[Reis \emph{et~al.}(2015)Reis, Jaeger, and Van~Hecke]{Reis2015}
	P.~M. Reis, H.~M. Jaeger and M.~Van~Hecke, \emph{Extreme Mechanics Letters},
	2015, \textbf{5}, 25--29\relax
	\mciteBstWouldAddEndPuncttrue
	\mciteSetBstMidEndSepPunct{\mcitedefaultmidpunct}
	{\mcitedefaultendpunct}{\mcitedefaultseppunct}\relax
	\EndOfBibitem
	\bibitem[Klein \emph{et~al.}(2007)Klein, Efrati, and Sharon]{Klein2007b}
	Y.~Klein, E.~Efrati and E.~Sharon, \emph{Science}, 2007, \textbf{315},
	1116--1120\relax
	\mciteBstWouldAddEndPuncttrue
	\mciteSetBstMidEndSepPunct{\mcitedefaultmidpunct}
	{\mcitedefaultendpunct}{\mcitedefaultseppunct}\relax
	\EndOfBibitem
	\bibitem[Holmes \emph{et~al.}(2011)Holmes, Roch{\'e}, Sinha, and
	Stone]{Holmes2011}
	D.~P. Holmes, M.~Roch{\'e}, T.~Sinha and H.~Stone, \emph{Soft Matter}, 2011,
	\textbf{7}, 5188\relax
	\mciteBstWouldAddEndPuncttrue
	\mciteSetBstMidEndSepPunct{\mcitedefaultmidpunct}
	{\mcitedefaultendpunct}{\mcitedefaultseppunct}\relax
	\EndOfBibitem
	\bibitem[Kim \emph{et~al.}(2012)Kim, Hanna, Byun, Santangelo, and
	Hayward]{Kim2012a}
	J.~Kim, J.~A. Hanna, M.~Byun, C.~D. Santangelo and R.~C. Hayward,
	\emph{Science}, 2012, \textbf{335}, 1201--1205\relax
	\mciteBstWouldAddEndPuncttrue
	\mciteSetBstMidEndSepPunct{\mcitedefaultmidpunct}
	{\mcitedefaultendpunct}{\mcitedefaultseppunct}\relax
	\EndOfBibitem
	\bibitem[Dervaux and Amar(2012)]{Dervaux2012}
	J.~Dervaux and M.~B. Amar, \emph{Annual Review of Condensed Matter Physics},
	2012, \textbf{3}, 311--332\relax
	\mciteBstWouldAddEndPuncttrue
	\mciteSetBstMidEndSepPunct{\mcitedefaultmidpunct}
	{\mcitedefaultendpunct}{\mcitedefaultseppunct}\relax
	\EndOfBibitem
	\bibitem[Pandey and Holmes(2013)]{Pandey2013}
	A.~Pandey and D.~P. Holmes, \emph{Soft Matter}, 2013, \textbf{9}, 5524\relax
	\mciteBstWouldAddEndPuncttrue
	\mciteSetBstMidEndSepPunct{\mcitedefaultmidpunct}
	{\mcitedefaultendpunct}{\mcitedefaultseppunct}\relax
	\EndOfBibitem
	\bibitem[Wu \emph{et~al.}(2013)Wu, Moshe, Greener, Therien-Aubin, Nie, Sharon,
	and Kumacheva]{Wu2013}
	Z.~L. Wu, M.~Moshe, J.~Greener, H.~Therien-Aubin, Z.~Nie, E.~Sharon and
	E.~Kumacheva, \emph{Nature Communications}, 2013, \textbf{4}, 1586--7\relax
	\mciteBstWouldAddEndPuncttrue
	\mciteSetBstMidEndSepPunct{\mcitedefaultmidpunct}
	{\mcitedefaultendpunct}{\mcitedefaultseppunct}\relax
	\EndOfBibitem
	\bibitem[Pezzulla \emph{et~al.}(2016)Pezzulla, Smith, Nardinocchi, and
	Holmes]{pezzulla16}
	M.~Pezzulla, G.~P. Smith, P.~Nardinocchi and D.~P. Holmes, \emph{Soft Matter},
	2016, \textbf{12}, 4435--4442\relax
	\mciteBstWouldAddEndPuncttrue
	\mciteSetBstMidEndSepPunct{\mcitedefaultmidpunct}
	{\mcitedefaultendpunct}{\mcitedefaultseppunct}\relax
	\EndOfBibitem
	\bibitem[Pezzulla \emph{et~al.}(2015)Pezzulla, Shillig, Nardinocchi, and
	Holmes]{pezzulla15}
	M.~Pezzulla, S.~A. Shillig, P.~Nardinocchi and D.~P. Holmes, \emph{Soft
		Matter}, 2015, \textbf{11}, 5812--5820\relax
	\mciteBstWouldAddEndPuncttrue
	\mciteSetBstMidEndSepPunct{\mcitedefaultmidpunct}
	{\mcitedefaultendpunct}{\mcitedefaultseppunct}\relax
	\EndOfBibitem
	\bibitem[Pezzulla \emph{et~al.}(2018)Pezzulla, Stoop, Steranka, Bade, and
	Holmes]{pezzulla18}
	M.~Pezzulla, N.~Stoop, M.~P. Steranka, A.~J. Bade and D.~P. Holmes,
	\emph{Physical Review Letters}, 2018, \textbf{120}, 048002\relax
	\mciteBstWouldAddEndPuncttrue
	\mciteSetBstMidEndSepPunct{\mcitedefaultmidpunct}
	{\mcitedefaultendpunct}{\mcitedefaultseppunct}\relax
	\EndOfBibitem
	\bibitem[Mora and Boudaoud(2006)]{Mora2006}
	T.~Mora and A.~Boudaoud, \emph{The European Physical Journal E}, 2006,
	\textbf{20}, 119--124\relax
	\mciteBstWouldAddEndPuncttrue
	\mciteSetBstMidEndSepPunct{\mcitedefaultmidpunct}
	{\mcitedefaultendpunct}{\mcitedefaultseppunct}\relax
	\EndOfBibitem
	\bibitem[DuPont \emph{et~al.}(2010)DuPont, Cates, Stroot, and
	Toomey]{DuPont2010}
	S.~J. DuPont, Jr, R.~S. Cates, P.~G. Stroot and R.~Toomey, \emph{Soft Matter},
	2010, \textbf{6}, 3876\relax
	\mciteBstWouldAddEndPuncttrue
	\mciteSetBstMidEndSepPunct{\mcitedefaultmidpunct}
	{\mcitedefaultendpunct}{\mcitedefaultseppunct}\relax
	\EndOfBibitem
	\bibitem[Barros \emph{et~al.}(2012)Barros, de~Azevedo, and
	Engelsberg]{Barros2012}
	W.~Barros, E.~N. de~Azevedo and M.~Engelsberg, \emph{Soft Matter}, 2012,
	\textbf{8}, 8511\relax
	\mciteBstWouldAddEndPuncttrue
	\mciteSetBstMidEndSepPunct{\mcitedefaultmidpunct}
	{\mcitedefaultendpunct}{\mcitedefaultseppunct}\relax
	\EndOfBibitem
	\bibitem[Efrati \emph{et~al.}(2009)Efrati, Sharon, and Kupferman]{efrati2009}
	E.~Efrati, E.~Sharon and R.~Kupferman, \emph{Journal of the Mechanics and
		Physics of Solids}, 2009, \textbf{57}, 762--775\relax
	\mciteBstWouldAddEndPuncttrue
	\mciteSetBstMidEndSepPunct{\mcitedefaultmidpunct}
	{\mcitedefaultendpunct}{\mcitedefaultseppunct}\relax
	\EndOfBibitem
	\bibitem[Dias \emph{et~al.}(2011)Dias, Hanna, and Santangelo]{dias2011}
	M.~A. Dias, J.~A. Hanna and C.~D. Santangelo, \emph{Physical Review E}, 2011,
	\textbf{84}, 036603\relax
	\mciteBstWouldAddEndPuncttrue
	\mciteSetBstMidEndSepPunct{\mcitedefaultmidpunct}
	{\mcitedefaultendpunct}{\mcitedefaultseppunct}\relax
	\EndOfBibitem
	\bibitem[van Rees \emph{et~al.}(2017)van Rees, Vouga, and
	Mahadevan]{vanRees2017}
	W.~M. van Rees, E.~Vouga and L.~Mahadevan, \emph{Proceedings of the National
		Academy of Sciences}, 2017,  201709025\relax
	\mciteBstWouldAddEndPuncttrue
	\mciteSetBstMidEndSepPunct{\mcitedefaultmidpunct}
	{\mcitedefaultendpunct}{\mcitedefaultseppunct}\relax
	\EndOfBibitem
	\bibitem[Domokos and Royce(1997)]{domokos97}
	P.~Domokos, G;~Holmes and B.~Royce, \emph{Journal of Nonlinear Science}, 1997,
	\textbf{7}, 281--314\relax
	\mciteBstWouldAddEndPuncttrue
	\mciteSetBstMidEndSepPunct{\mcitedefaultmidpunct}
	{\mcitedefaultendpunct}{\mcitedefaultseppunct}\relax
	\EndOfBibitem
	\bibitem[Holmes \emph{et~al.}(1999)Holmes, Domokos, Schmitt, and
	Szeber{\'e}nyi]{holmes99}
	P.~Holmes, G.~Domokos, J.~Schmitt and I.~Szeber{\'e}nyi, \emph{Computer Methods
		in Applied Mechanics and Engineering}, 1999, \textbf{170}, 175--207\relax
	\mciteBstWouldAddEndPuncttrue
	\mciteSetBstMidEndSepPunct{\mcitedefaultmidpunct}
	{\mcitedefaultendpunct}{\mcitedefaultseppunct}\relax
	\EndOfBibitem
	\bibitem[Roman and Pocheau(1999)]{roman99}
	B.~Roman and A.~Pocheau, \emph{EPL (Europhysics Letters)}, 1999, \textbf{46},
	602--608\relax
	\mciteBstWouldAddEndPuncttrue
	\mciteSetBstMidEndSepPunct{\mcitedefaultmidpunct}
	{\mcitedefaultendpunct}{\mcitedefaultseppunct}\relax
	\EndOfBibitem
	\bibitem[Roman and Pocheau(2002)]{roman02}
	B.~Roman and A.~Pocheau, \emph{Journal of the Mechanics and Physics of Solids},
	2002, \textbf{50}, 2379--2401\relax
	\mciteBstWouldAddEndPuncttrue
	\mciteSetBstMidEndSepPunct{\mcitedefaultmidpunct}
	{\mcitedefaultendpunct}{\mcitedefaultseppunct}\relax
	\EndOfBibitem
	\bibitem[Pocheau and Roman(2004)]{pocheau04}
	A.~Pocheau and B.~Roman, \emph{Physica D: Nonlinear Phenomena}, 2004,
	\textbf{192}, 161--186\relax
	\mciteBstWouldAddEndPuncttrue
	\mciteSetBstMidEndSepPunct{\mcitedefaultmidpunct}
	{\mcitedefaultendpunct}{\mcitedefaultseppunct}\relax
	\EndOfBibitem
	\bibitem[Ben~Amar and Pomeau(1997)]{BenAmar1997}
	M.~Ben~Amar and Y.~Pomeau, \emph{Proceedings of the Royal Society of London A:
		Mathematical, Physical and Engineering Sciences}, 1997, \textbf{453},
	729--755\relax
	\mciteBstWouldAddEndPuncttrue
	\mciteSetBstMidEndSepPunct{\mcitedefaultmidpunct}
	{\mcitedefaultendpunct}{\mcitedefaultseppunct}\relax
	\EndOfBibitem
	\bibitem[Cerda and Mahadevan(1998)]{Cerda1998}
	E.~Cerda and L.~Mahadevan, \emph{Physical Review Letters}, 1998, \textbf{80},
	2358\relax
	\mciteBstWouldAddEndPuncttrue
	\mciteSetBstMidEndSepPunct{\mcitedefaultmidpunct}
	{\mcitedefaultendpunct}{\mcitedefaultseppunct}\relax
	\EndOfBibitem
	\bibitem[Cha{\"\i}eb \emph{et~al.}(1998)Cha{\"\i}eb, Melo, and
	G{\'e}minard]{Chaieb1998}
	S.~Cha{\"\i}eb, F.~Melo and J.-C. G{\'e}minard, \emph{Physical Review Letters},
	1998, \textbf{80}, 2354\relax
	\mciteBstWouldAddEndPuncttrue
	\mciteSetBstMidEndSepPunct{\mcitedefaultmidpunct}
	{\mcitedefaultendpunct}{\mcitedefaultseppunct}\relax
	\EndOfBibitem
	\bibitem[Cerda \emph{et~al.}(1999)Cerda, Chaieb, Melo, and
	Mahadevan]{Cerda1999}
	E.~Cerda, S.~Chaieb, F.~Melo and L.~Mahadevan, \emph{Nature}, 1999,
	\textbf{401}, 46--49\relax
	\mciteBstWouldAddEndPuncttrue
	\mciteSetBstMidEndSepPunct{\mcitedefaultmidpunct}
	{\mcitedefaultendpunct}{\mcitedefaultseppunct}\relax
	\EndOfBibitem
	\bibitem[Chaieb(2000)]{Chaieb2000}
	S.~Chaieb, \emph{Journal of the Mechanics and Physics of Solids}, 2000,
	\textbf{48}, 565--579\relax
	\mciteBstWouldAddEndPuncttrue
	\mciteSetBstMidEndSepPunct{\mcitedefaultmidpunct}
	{\mcitedefaultendpunct}{\mcitedefaultseppunct}\relax
	\EndOfBibitem
	\bibitem[Cerda and Mahadevan(2005)]{cerda05}
	E.~Cerda and L.~Mahadevan, \emph{Proceedings of the Royal Society of London A:
		Mathematical, Physical and Engineering Sciences}, 2005, \textbf{461},
	671--700\relax
	\mciteBstWouldAddEndPuncttrue
	\mciteSetBstMidEndSepPunct{\mcitedefaultmidpunct}
	{\mcitedefaultendpunct}{\mcitedefaultseppunct}\relax
	\EndOfBibitem
	\bibitem[Hure \emph{et~al.}(2012)Hure, Roman, and Bico]{hure12}
	J.~Hure, B.~Roman and J.~Bico, \emph{Physical Review Letters}, 2012,
	\textbf{109}, 054302\relax
	\mciteBstWouldAddEndPuncttrue
	\mciteSetBstMidEndSepPunct{\mcitedefaultmidpunct}
	{\mcitedefaultendpunct}{\mcitedefaultseppunct}\relax
	\EndOfBibitem
	\bibitem[Paulsen \emph{et~al.}(2015)Paulsen, D{\'e}mery, Santangelo, Russell,
	Davidovitch, and Menon]{Paulsen2015}
	J.~D. Paulsen, V.~D{\'e}mery, C.~D. Santangelo, T.~P. Russell, B.~Davidovitch
	and N.~Menon, \emph{Nature Materials}, 2015, \textbf{14}, 1206\relax
	\mciteBstWouldAddEndPuncttrue
	\mciteSetBstMidEndSepPunct{\mcitedefaultmidpunct}
	{\mcitedefaultendpunct}{\mcitedefaultseppunct}\relax
	\EndOfBibitem
	\bibitem[Vaziri and Mahadevan(2008)]{vaziri08}
	A.~Vaziri and L.~Mahadevan, \emph{Proceedings of the National Academy of
		Sciences}, 2008, \textbf{105}, 7913--7918\relax
	\mciteBstWouldAddEndPuncttrue
	\mciteSetBstMidEndSepPunct{\mcitedefaultmidpunct}
	{\mcitedefaultendpunct}{\mcitedefaultseppunct}\relax
	\EndOfBibitem
	\bibitem[Nasto \emph{et~al.}(2013)Nasto, Ajdari, Lazarus, Vaziri, and
	Reis]{nasto13}
	A.~Nasto, A.~Ajdari, A.~Lazarus, A.~Vaziri and P.~M. Reis, \emph{Soft Matter},
	2013, \textbf{9}, 6796--9\relax
	\mciteBstWouldAddEndPuncttrue
	\mciteSetBstMidEndSepPunct{\mcitedefaultmidpunct}
	{\mcitedefaultendpunct}{\mcitedefaultseppunct}\relax
	\EndOfBibitem
	\bibitem[Vella and Davidovitch(2018)]{Vella2018}
	D.~Vella and B.~Davidovitch, \emph{arXiv preprint arXiv:1804.03341}, 2018\relax
	\mciteBstWouldAddEndPuncttrue
	\mciteSetBstMidEndSepPunct{\mcitedefaultmidpunct}
	{\mcitedefaultendpunct}{\mcitedefaultseppunct}\relax
	\EndOfBibitem
	\bibitem[Paulsen(2018)]{Paulsen2018}
	J.~D. Paulsen, \emph{arXiv preprint arXiv:1804.07425}, 2018\relax
	\mciteBstWouldAddEndPuncttrue
	\mciteSetBstMidEndSepPunct{\mcitedefaultmidpunct}
	{\mcitedefaultendpunct}{\mcitedefaultseppunct}\relax
	\EndOfBibitem
	\bibitem[Mahadevan \emph{et~al.}(2007)Mahadevan, Vaziri, and Das]{mahadevan07}
	L.~Mahadevan, A.~Vaziri and M.~Das, \emph{EPL (Europhysics Letters)}, 2007,
	\textbf{77}, 40003\relax
	\mciteBstWouldAddEndPuncttrue
	\mciteSetBstMidEndSepPunct{\mcitedefaultmidpunct}
	{\mcitedefaultendpunct}{\mcitedefaultseppunct}\relax
	\EndOfBibitem
	\bibitem[Jiang \emph{et~al.}(2018)Jiang, Pezzulla, Shao, Ghosh, and
	Holmes]{jiang18}
	X.~Jiang, M.~Pezzulla, H.~Shao, T.~K. Ghosh and D.~P. Holmes, \emph{EPL
		(Europhysics Letters)}, 2018, \textbf{122}, 64003\relax
	\mciteBstWouldAddEndPuncttrue
	\mciteSetBstMidEndSepPunct{\mcitedefaultmidpunct}
	{\mcitedefaultendpunct}{\mcitedefaultseppunct}\relax
	\EndOfBibitem
	\bibitem[Klein \emph{et~al.}(2007)Klein, Efrati, and Sharon]{klein07}
	Y.~Klein, E.~Efrati and E.~Sharon, \emph{Science}, 2007, \textbf{315},
	1116--1120\relax
	\mciteBstWouldAddEndPuncttrue
	\mciteSetBstMidEndSepPunct{\mcitedefaultmidpunct}
	{\mcitedefaultendpunct}{\mcitedefaultseppunct}\relax
	\EndOfBibitem
	\bibitem[Pezzulla \emph{et~al.}(2017)Pezzulla, Stoop, Jiang, and
	Holmes]{pezzulla17}
	M.~Pezzulla, N.~Stoop, X.~Jiang and D.~P. Holmes, \emph{Proceedings of the
		Royal Society of London A: Mathematical, Physical and Engineering Sciences},
	2017, \textbf{473}, 20170087\relax
	\mciteBstWouldAddEndPuncttrue
	\mciteSetBstMidEndSepPunct{\mcitedefaultmidpunct}
	{\mcitedefaultendpunct}{\mcitedefaultseppunct}\relax
	\EndOfBibitem
	\bibitem[Lee \emph{et~al.}()Lee, Brun, Marthelot, Nature, and {2016}]{lee16}
	A.~Lee, P.~T. Brun, J.~Marthelot, G.~B. Nature and {2016}, \emph{Nature
		Communications}\relax
	\mciteBstWouldAddEndPuncttrue
	\mciteSetBstMidEndSepPunct{\mcitedefaultmidpunct}
	{\mcitedefaultendpunct}{\mcitedefaultseppunct}\relax
	\EndOfBibitem
	\bibitem[{Marthelot} \emph{et~al.}(2017){Marthelot}, {Brun}, {Lopez Jimenez},
	and {Reis}]{marthelot17}
	J.~{Marthelot}, P.-T. {Brun}, F.~{Lopez Jimenez} and P.~M. {Reis}, APS Meeting
	Abstracts, 2017, p. C15.010\relax
	\mciteBstWouldAddEndPuncttrue
	\mciteSetBstMidEndSepPunct{\mcitedefaultmidpunct}
	{\mcitedefaultendpunct}{\mcitedefaultseppunct}\relax
	\EndOfBibitem
	\bibitem[Lucantonio \emph{et~al.}(2014)Lucantonio, Nardinocchi, and
	Pezzulla]{lucantonio14}
	A.~Lucantonio, P.~Nardinocchi and M.~Pezzulla, \emph{Proceedings of the Royal
		Society of London A: Mathematical, Physical and Engineering Sciences}, 2014,
	\textbf{470}, 20140467--20140467\relax
	\mciteBstWouldAddEndPuncttrue
	\mciteSetBstMidEndSepPunct{\mcitedefaultmidpunct}
	{\mcitedefaultendpunct}{\mcitedefaultseppunct}\relax
	\EndOfBibitem

\end{thebibliography}
\end{document}